\documentstyle[prb,aps,eqsecnum,multicol,epsf]{revtex}

\newcommand{\bleq}{\ifpreprintsty
                   \else
                   \end{multicols}\vspace*{-3.5ex}{\tiny
                   \noindent\begin{tabular}[t]{c|}
                   \parbox{0.493\hsize}{~} \\ \hline \end{tabular}}
                   \fi} 
\newcommand{\eleq}{\ifpreprintsty
                  \else
                   {\tiny\hspace*{\fill}\begin{tabular}[t]{|c}\hline
                    \parbox{0.49\hsize}{~} \\
                    \end{tabular}}\vspace*{-2.5ex}\begin{multicols}{2}
                    \fi}
\newcommand{\bcols}{\ifpreprintsty\else\begin{multicols}{2}\fi}
\newcommand{\ecols}{\ifpreprintsty\else\end{multicols}\fi}

\begin{document}
\draft

\title{Spin waves in the spiral phase of a doped
                   antiferromagnet: a strong-coupling approach}    
\author{N. Dupuis }
\address{ Laboratoire de Physique des Solides, Associ\'e au CNRS,
Universit\'e Paris-Sud, 91405 Orsay, France } 
\date{May 3, 2001}
\maketitle

\begin{abstract}  
We study spin fluctuations in the spiral phase of the two-dimensional
Hubbard model at low doping on the basis of the spin-particle-hole
coherent-state path integral. In the strong correlation limit, we
obtain an analytical expression of the spin-wave excitations over the
entire Brillouin zone except in the vicinity of ${\bf q}=0$.
We discuss the validity of the Hartree-Fock and random-phase
approximations in the strong-coupling limit, and 
compare our results with previous numerical and analytical
calculations. Although the spiral phase is unstable, as shown by a
negative mean-field compressibility and the presence of imaginary
spin-fluctuation modes, we expect the short-wavelength fluctuation
modes (with real energies) to survive in the actual ground-state of
the system. 
\end{abstract}


\bcols 


\section{Introduction}

Despite a lot of theoretical efforts, there is still no satisfying
description of the ground-state and the low-lying excitations of a
doped two-dimensional (2D) antiferromagnet. One of the simplest
(realistic) models describing such a system is the Hubbard model. In
the strong correlation limit, it reduces to the Heisenberg model at
half-filling. For a 2D square lattice, the ground-state is known to be
antiferromagnetic, and the low-lying excitations (spin-waves) are well
understood. However, away from half-filling, the coexistence of local
moments and itinerant charge carriers raises difficulties that have
not been overcome so far. 

In the framework of the Heisenberg model, perturbation theory around a
broken-spin-symmetry ground-state (from linear spin-wave analysis to
renormalization-group approach) has proven to be very
successful. Formally, this semiclassical approach corresponds to a
$1/S$ expansion, where $S$ is the size of the localized spins. It is
very natural to follow the same line of approach in itinerant
spin-$\frac{1}{2}$ fermion systems by promoting the localized moments
from spin-$\frac{1}{2}$ to spin-$S$ (with $S\gg 1$). The spin degrees
of freedom are then treated semiclassically, whereas the itinerant
fermions are considered quantum mechanically. 

The semiclassical approach is realized for instance in the
Hartree-Fock or random-phase (RPA) approximations. For the Hubbard 
model, the (homogeneous) Hartree-Fock theory
\cite{Schulz90,Singh91,Arrigoni91,Dzierzawa92,Zhou95,Zhou94,Chubukov95,Brenig96,Kampf96,Arrigoni00}
and the slave-boson  mean-field theory\cite{Arrigoni91,Fresard91}
predict a spiral magnetic order at strong coupling in agreement with
mean-field (semiclassical) theories of the $t$-$J$ model
\cite{Jayaprakash89,Yoshioka89,Kane90,Auerbach91,Gan91} as first
shown by Schraiman and Siggia.\cite{Schraiman89}  This
(homogeneous) spiral phase turns out to be unstable, as shown by a
negative compressibility. The instability also manifests itself by the
presence of imaginary spin-wave modes. 
\cite{Zhou95,Zhou94,Brenig96,Arrigoni00} Short-range spiral
order,\cite{Schraiman89}, phase separation, coexisting
spin- and charge-density waves (domain-wall formation),
\cite{Dombre89,Zhou95,Zhou94,Brenig96} and formation of local spin
polarons\cite{Schrieffer88,Singh90,Auerbach91a} have been proposed as
a possible alternative to the spiral phase. Although phase separation
is likely to be suppressed by   
long-range Coulomb interaction, it is not clear whether the latter can
stabilize the homogeneous spiral magnetic order. \cite{Zhou95,Zhou94}
In any case, we expect the short-range spiral order to survive in the
actual ground-state of the system. The study of (short-wavelength)
spin-wave modes around the mean-field spiral order can then be seen as
a way to access this short-range order, even though the mean-field
theory incorrectly predicts a long-range spiral order. On the
experimental side,  strong antiferromagnetic fluctuations have been
observed in the normal phase of high-$T_c$ copper oxides. In
particular, inelastic neutron scattering data have revealed noticeable
incommensurate fluctuations at finite doping. \cite{Exp}

In this paper, we reconsider the semiclassical limit of the 2D Hubbard
model on the basis of the spin-particle-hole coherent-state path
integral. \cite{Dupuis00,ND} The latter has been designed to study
strongly correlated fermion systems where itinerant charge carriers
interact with local moments. As shown below, it provides a very
natural framework to derive the spin-wave modes around the mean-field
spiral order. 

The outline of the paper is as follows. In Sec.~\ref{sec:mft}, we
introduce the effective action of the Hubbard model based on the
spin-particle-hole coherent-state path integral. After performing a
saddle-point approximation on the spin variables corresponding to a
(homogeneous) spiral order, we calculate the hole Green's function and
the free energy within a $t/U$ expansion. We discuss the reason why
the Hartree-Fock theory turns out to be correct in the strong correlation limit
at low doping. In Sec.~\ref{sec:swm}, we compute the spin-wave
excitations. We point out that a correct description of the three
Goldstone modes of the spiral phase (at ${\bf q}=0,\pm {\bf Q}$, where
${\bf Q}$ is the wave-vector of the spin modulation) follows from a
consistent calculation of vertex corrections and self-energy terms in
the fermion Green's functions. Finally, we obtain an analytical
expression of the spin-wave excitations valid over the entire
Brillouin zone, except in the vicinity of ${\bf q}=0$. We compare our
analytical result with previous numerical
\cite{Zhou95,Zhou94,Brenig96,Kampf96} and analytical
calculations.\cite{Arrigoni00}

\section{Mean-field theory}
\label{sec:mft}

We consider a bipartite 2D lattice with $N$ sites. The Hubbard model
is defined by 
\begin{equation}
\hat H=-t \sum_{\langle{\bf r},{\bf r}'\rangle,\sigma} (\hat c^\dagger_{{\bf r}
\sigma}\hat c_{{\bf r}'\sigma} +{\rm h.c.} ) + U \sum_{\bf r} \hat n_{{\bf r}
\uparrow} \hat n_{{\bf r}\downarrow} ,
\label{Ham}
\end{equation}
where $\hat c_{{\bf r}\sigma}$ is a fermionic operator for a $\sigma$-spin 
particle at site $\bf r$ ($\sigma=\uparrow,\downarrow$), $\hat n_{{\bf
r}\sigma}=\hat c^\dagger_{{\bf r} \sigma}\hat c_{{\bf r}\sigma}$, and
$\langle {\bf r},{\bf r}'\rangle$ denotes nearest neighbors. We denote
by $\mu$ the chemical potential, $\beta$ the inverse temperature, and
$n=1-x$ the mean number of particles per site. We consider only hole
doping ($x\geq 0$) and set $\hbar=k_B=1$ throughout the paper. All
results are obtained in the zero-temperature limit ($T\to 0$). 

The spin-particle-hole coherent-state path integral formulation of the
Hubbard model was derived in Refs.~\onlinecite{Dupuis00,ND}. This
approach is based on the introduction of spin-particle-hole coherent
states which generalize the spin-$\frac{1}{2}$ coherent states by
allowing the creation of a hole or an additional
particle. In the strong-coupling limit $U\gg t$, the effective action
$S[\gamma^*,\gamma;{\bf\Omega}]$ of the Hubbard model is given by 
\bleq
\begin{eqnarray}   
S[\gamma^*,\gamma;{\bf\Omega}] &=& \sum_{\bf r} \int d\tau
\gamma^*_{{\bf r}\uparrow}  
(\partial_\tau-\mu+A^0_{\bf r}) \gamma_{{\bf r}\uparrow} 
-\sum_{{\bf r},{\bf r}'} \int d\tau \gamma^*_{{\bf r}\uparrow} 
\hat t_{{\bf r}\uparrow,{\bf r}'\uparrow} \gamma_{{\bf r}'\uparrow} 
\nonumber \\ && 
+\sum_{\bf r} \int d\tau \gamma^*_{{\bf r}\downarrow} 
(\partial_\tau-\mu+U-A^0_{\bf r}) \gamma_{{\bf r}\downarrow} 
-\sum_{{\bf r},{\bf r}'} \int d\tau \gamma^*_{{\bf r}\downarrow} 
\hat t_{{\bf r}\downarrow,{\bf r}'\downarrow} \gamma_{{\bf r}'\downarrow} 
\nonumber \\ && 
-\sum_{{\bf r},{\bf r}'} \int d\tau \Bigl( \gamma^*_{{\bf r}\uparrow} 
\hat t_{{\bf r}\uparrow,{\bf r}'\downarrow} \gamma_{{\bf
r}'\downarrow} +{\rm c.c.} \Bigr)  
\nonumber  \\ && 
+\sum_{\bf r} \int d\tau_1 d\tau_2  d\tau_3 d\tau_4
\Gamma^{\rm
II}_{\uparrow\downarrow,\uparrow\downarrow}(\tau_1,\tau_2;\tau_3,\tau_4)  
\gamma^*_{{\bf r}\uparrow}(\tau_1)\gamma^*_{{\bf r}\downarrow}(\tau_2) 
\gamma_{{\bf r}\downarrow}(\tau_4)\gamma_{{\bf r} \uparrow}(\tau_3) .
\label{action1}
\end{eqnarray} 
\eleq
${\bf\Omega}$ is a unit vector field which gives the direction of the
local moments at the singly-occupied sites. 
$\gamma_\uparrow$ and $\gamma_\downarrow$ are Grassmann variables
describing particles propagating in the lower (LHB) and upper (UHB)
Hubbard bands, respectively. $A^0_{\bf r}=\langle {\bf\Omega}_{\bf r}
|\dot{\bf\Omega}_{\bf r}\rangle$ is a Berry phase term ($|\dot
{\bf\Omega}_{\bf r}\rangle=\partial_\tau| {\bf\Omega}_{\bf r}\rangle$). The
intersite hopping matrix $\hat t_{{\bf r}{\bf 
r}'}=R^\dagger_{\bf r} t_{{\bf r}{\bf r}'}R_{{\bf r}'}$ 
depends on ${\bf\Omega}$ {\it via} the SU(2)/U(1) matrix
\begin{equation}
R_{\bf r} = e^{-\frac{i}{2}\varphi_{\bf r} \sigma_z}  
        e^{-\frac{i}{2}\theta_{\bf r} \sigma_y} 
        e^{-\frac{i}{2}\psi_{\bf r} \sigma_z}
\label{R1}  
\end{equation}
which rotates the unit vector $\hat {\bf z}$ to
${\bf\Omega}_{\bf r}={\bf\Omega}(\theta_{\bf r},\varphi_{\bf r})$. Here
$\theta_{\bf r}$ and $\varphi_{\bf r}$ are the polar angles
determining the direction of ${\bf\Omega}_{\bf r}$. The choice of
$\psi_{\bf r}$ is free and corresponds to a ``gauge'' freedom. In the
Hubbard model, $t_{{\bf rr}'}=t$ if ${\bf r}$ and ${\bf r}'$ are
nearest neighbors and vanishes otherwise. The quartic term of the
action (\ref{action1}) is determined by the two-particle atomic vertex
$\Gamma^{\rm II}$.  

In this section, we make a saddle-point approximation on the spin
variables ${\bf\Omega}_{\bf r}$. As discussed in Ref.~\onlinecite{ND}, this
approach can be 
justified by taking a ``large-$S$'' semiclassical limit, which
consists in promoting the spin-$\frac{1}{2}$ coherent states describing singly
occupied sites to spin-$S$ coherent states. 
In the limit $S\to\infty$, the Berry phase term suppresses quantum
fluctuations of ${\bf\Omega}$. The spin variables become classical and
do not fluctuate at zero temperature. 

A broken-symmetry ground-state corresponding to a (coplanar) spiral order
is defined by the ``classical'' configuration
\begin{equation}
\theta^{\rm cl}_{\bf r} = \frac{\pi}{2}, \,\,\, \,\,\,\,\,\,
\varphi^{\rm cl}_{\bf r} = {\bf Q} \cdot {\bf r} .
\label{cl1}
\end{equation}
We consider only the diagonal spiral phase (${\bf Q}=(Q,Q)$), which is
known to be the most stable one in the strong-coupling limit.\cite{Schulz90} 
Note that the antiferromagnetic (${\bf Q}=(\pi,\pi)$) and
ferromagnetic (${\bf Q}=0$) phases are special cases of the spiral
order defined by Eq.~(\ref{cl1}).
In the gauge $\psi_{\bf r}=0$, the hopping matrix $\hat
t^{\rm cl}_{{\bf r}{\bf r}'}$ depends only on the difference ${\bf
r}-{\bf r}'$: 
\begin{equation} 
\hat t^{\rm cl}_{{\bf r}{\bf r}'} =  t_{{\bf r}{\bf r}'} 
e^{-\frac{i}{2}{\bf Q}\cdot ({\bf r}-{\bf r}')\sigma_x} .
\end{equation}
This yields the following saddle-point action:
\bleq
\begin{eqnarray}
S_{\rm cl}[\gamma^*,\gamma] &=& - \sum_{{\bf k},\omega,\sigma}
\gamma^*_{{\bf k}\sigma}(i\omega) 
( i\omega+\mu-U\delta_{\sigma\downarrow} +2t\cos\frac{Q}{2} \cos k_\nu) 
\gamma_{{\bf k}\sigma}(i\omega) 
+ \sum_{{\bf k},\omega,\sigma} \gamma^*_{{\bf k}\sigma}(i\omega) 
2t \sin\frac{Q}{2} \sin k_\nu \gamma_{{\bf k}\bar\sigma}(i\omega)
\nonumber \\ && + \frac{1}{N\beta}
\sum_{{\bf k},{\bf k}',{\bf q}} \sum_{\omega_1,\omega_2,\omega_3} 
\Gamma^{\rm II}_{\uparrow\downarrow,\uparrow\downarrow}
(i\omega_1,i\omega_2;i\omega_3 (i\omega_4))  
\gamma^*_{{\bf k}+{\bf q}\uparrow}(i\omega_1)
\gamma^*_{{\bf k}'-{\bf q}\downarrow}(i\omega_2)
\gamma_{{\bf k}'\downarrow}(i\omega_4)
\gamma_{{\bf k}\uparrow}(i\omega_3), 
\label{action2}
\end{eqnarray}
\eleq
where $\gamma_{{\bf k}\sigma}$ is the Fourier transform of
$\gamma_{{\bf r}\sigma}$, and 
${\bf k}$ runs over the entire Brillouin zone. 
$\Gamma^{\rm II}_{\uparrow\downarrow,\uparrow\downarrow}
(i\omega_1,i\omega_2;\omega_3 (i\omega_4))=i\omega_2-i\omega_3-U$,
where 
$\omega_4=\omega_1+\omega_2-\omega_3$ is fixed by energy
conservation. In Eq.~(\ref{action2}) and below, there is an implicit
sum over $\nu=x,y$. Without the quartic term, the action
(\ref{action2}) is similar to that obtained in
Refs.~\onlinecite{Zhou95,Zhou94} within a large-$U$ Hartree-Fock
approximation where the SU(2) spin-rotation invariance is maintained
by introducing a fluctuating spin-quantization axis in the functional
integral. We show below that the quartic term can indeed be neglected
in the {\it low-doping} limit. 

For later convenience, we  introduce the parameter 
\begin{equation}
p=-\cos \frac{Q}{2} \geq 0
\end{equation}
which determines the spiral pitch. By changing $Q$ in $Q+2\pi$, it is
always possible to choose the sign of $p$ positive. $p$ varies from 0
in the antiferromagnetic phase to 1 in the ferromagnetic phase. 

\begin{figure}
\epsfysize 5 cm
\epsffile[30 410 280 570]{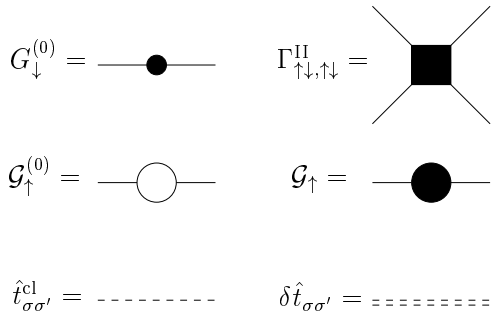}
\epsfysize 8 cm
\caption{ 
Definition of the various symbols appearing in the Feynman
diagrams. } 
\label{fig:def}
\end{figure}

\begin{figure}
\epsfysize 7 cm
\epsffile[170 200 450 580]{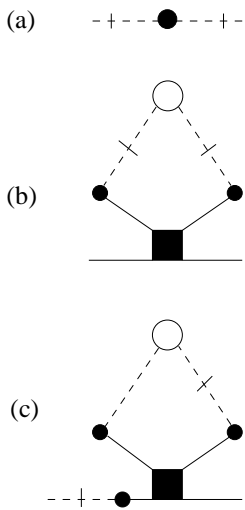}
\caption{
Diagrammatic representation of the LHB self-energy at order
$O(t/U)$.  Slashed dashed lines indicate interband transitions ($\hat
t^{\rm cl}_{\sigma\bar\sigma}$). }
\label{fig:SE}
\end{figure}

We are now in a position to apply the strong-coupling perturbation
theory described in Ref.~\onlinecite{ND}. We first consider the LHB
Green's function. To zeroth order in $t/U$, $\hat t^{\rm
cl}_{\sigma\bar\sigma}$ is ignored, since it corresponds to interband
transitions. As shown in Ref.~\onlinecite{ND}, $\Gamma^{\rm II}$ does
not affect the LHB Green's function to this order. We thus obtain
\begin{equation}
{\cal G}^{(0)}_\uparrow({\bf k},i\omega) 
= (i\omega+\mu-2tp \cos k_\nu)^{-1}
\end{equation}
as the propagator of the $\gamma_\uparrow$ field. 
Corrections to ${\cal G}^{(0)}_\uparrow$ are taken into account by
introducing a self-energy $\Sigma$, i.e. ${\cal
G}^{-1}_\uparrow={\cal G}^{(0)-1}_\uparrow -\Sigma$. To order
$O(t/U)$, there are three contributions to $\Sigma$ shown in
Fig.~\ref{fig:SE}.\cite{ND} [The symbols used in the Feynman diagrams are
defined in Fig.~\ref{fig:def}.] The last contribution (Fig.~\ref{fig:SE}c)
vanishes due 
to the sum over the internal momentum in the loop. We show at the end
of this section that the second contribution (Fig.~\ref{fig:SE}b) can be
neglected in the low-doping limit. The first contribution
(Fig.~\ref{fig:SE}a) gives
\begin{eqnarray}
\Sigma({\bf k},i\omega) &=& 4t^2(1-p^2)(\sin k_\nu)^2
G^{(0)}_\downarrow(i\omega) \nonumber \\ 
&\simeq&  -J(1-p^2)(\sin k_\nu)^2, 
\label{Sig1}
\end{eqnarray}
where $J=4t^2/U$. The last line of Eq.~(\ref{Sig1}) is obtained by
approximating the UHB atomic Green's function
$G^{(0)}_\downarrow(i\omega)=(i\omega+\mu-U)^{-1}$ by $-1/U$. We
therefore obtain the following expression for the propagator of the
$\gamma$ field:
\begin{eqnarray}
&& {\cal G}_\uparrow({\bf k},i\omega) = (i\omega-\epsilon_{\bf
k})^{-1},  
\label{Gup} \\ &&
\epsilon_{\bf k} = -\mu + 2tp \cos k_\nu-J(1-p^2)(\sin k_\nu)^2 .
\label{disp1}
\end{eqnarray}
It should be noted that ${\cal G}_\uparrow$ is not the LHB Green's
function of the original fermions (i.e. the $c$ field in
Eq.~(\ref{Ham})). The latter is given by (see Eq.~(3.32) in
Ref.~\onlinecite{ND}) 
\begin{eqnarray}
{\cal G}^{\rm LHB}_\sigma({\bf r}\tau,{\bf r}'\tau') &=& (R_{\bf
r}(\tau))_{\sigma\uparrow} {\cal G}_\uparrow ({\bf r}\tau,{\bf
r}'\tau')  (R_{\bf r}'(\tau'))_{\uparrow\sigma} \nonumber \\
&=& \frac{e^{-\frac{i}{2}\sigma {\bf Q}\cdot ({\bf r}-{\bf r}')}}{2} 
{\cal G}_\uparrow ({\bf r}\tau,{\bf r}'\tau').
\end{eqnarray}
In Fourier space, this gives
\begin{equation}
{\cal G}^{\rm LHB}_\sigma({\bf k},i\omega) = \frac{1}{2}{\cal
G}_\uparrow ({\bf k}+\frac{\sigma}{2}{\bf Q},i\omega) .
\label{disp1bis}
\end{equation}
The quasi-particle pole has a residue equal to $1/2$, as expected since
${\cal G}^{\rm LHB}_\sigma$ is the Green's function projected onto the
LHB. Moreover, the dispersion law of the original fermions is shifted
by $\pm {\bf Q}/2$ with respect to that of the $\gamma_\uparrow$
particles. 

Eqs.~(\ref{disp1}) and (\ref{disp1bis}) agrees with the large-$U$
expansion of the Hartree-Fock result. There are two contributions to
the energy $\epsilon_{\bf k}$ of the particles in the 
LHB. The first is due to inter-sublattice hopping processes and gives a
band of width $8tp$. The second comes from intra-sublattice hopping
processes which occur {\it via} virtual transitions to the UHB. The
resulting bandwidth is of order $J(1-p^2)$. Anticipating that $p\simeq
(U/2t)x$ [Eq.~(\ref{p1})], inter- (intra)-sublattice hopping processes
dominate when $J(1-p^2)\ll tp$, i.e $t^2/U^2\ll x$ ($tp\ll J(1-p^2)$,
i.e. $x\ll t^2/U^2$).  From Eq.~(\ref{disp1}), we readily see the
instability of the antiferromagnetic phase against the formation of a
spiral phase in the presence of holes. When $p$ is finite, the kinetic
energy gain due to inter-sublattice hopping (or order $p$) dominates
over the energy loss due to intra-sublattice hopping (of order $p^2$), thus
stabilizing a non-linear spin order. [Note that the energy loss due to
intra-sublattice hopping can also be seen as a loss of exchange energy.]

\begin{figure}
\epsfysize 13 cm
\epsffile[55 240 310 620]{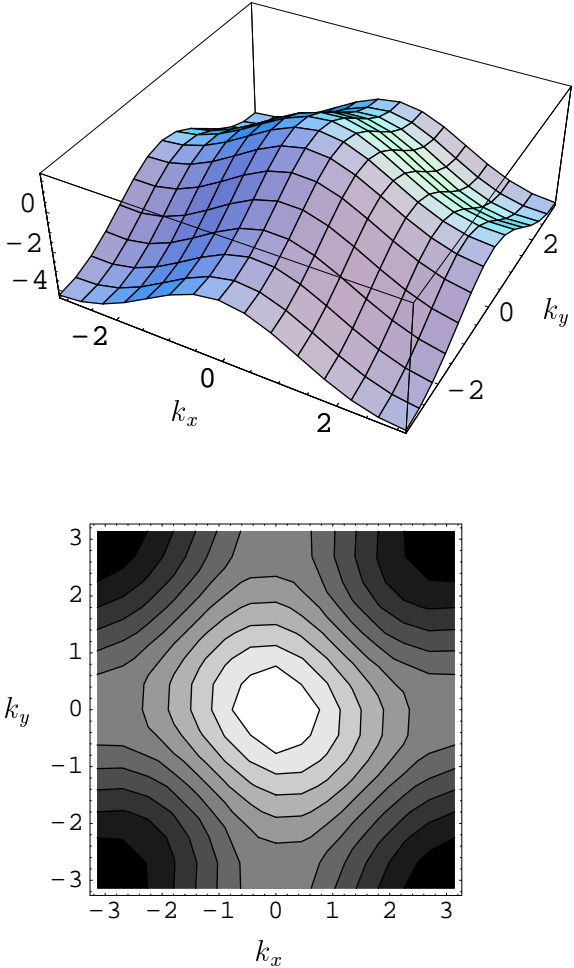}
\caption{
Dispersion law $\epsilon_{\bf k}$ in the LHB for $t^2/U^2\ll x$
($U/t=10$ and $x=0.15$). }
\label{fig:FS1}
\end{figure}

\begin{figure}
\epsfysize 13 cm
\epsffile[55 235 290 620]{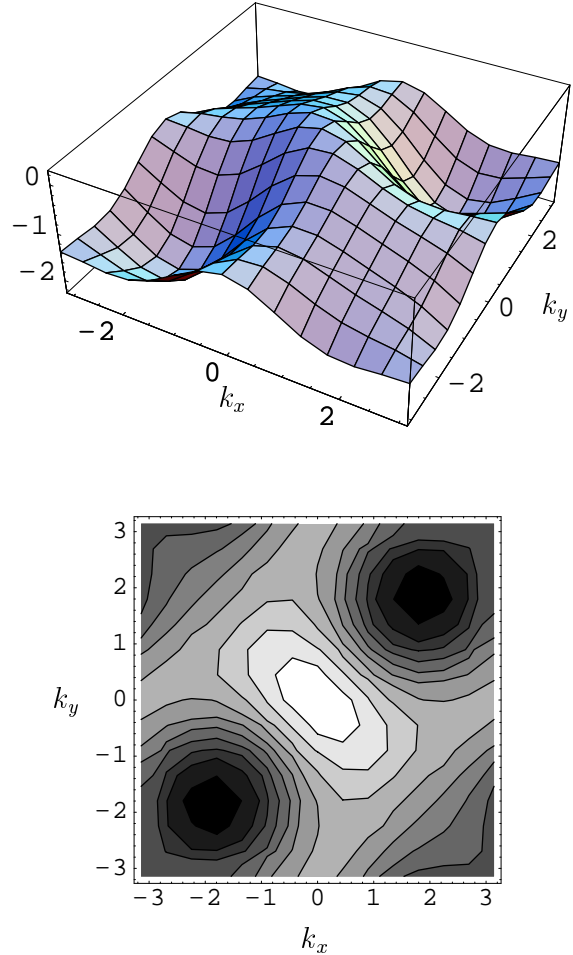}
\caption{
Dispersion law $\epsilon_{\bf k}$ in the LHB for $x\sim t^2/U^2$
($U/t=10$ and $x=0.04$). }
\label{fig:FS2}
\end{figure}

Since $p\geq 0$, the top of the LHB (for the $\gamma$ particles) is
located in the 
middle of the Brillouin zone (${\bf k}=0$). In the vicinity of ${\bf
k}=0$, the dispersion law (\ref{disp1}) can be approximated as
\begin{equation}
\epsilon_{\bf k} \simeq -\mu +4tp -tpk^2-J(1-p^2)(k_x+k_y)^2 .
\end{equation}
Introducing the momentum ${\bf K}$ defined by
\begin{equation}
K_x = \frac{k_x+k_y}{\sqrt{2}} , \,\,\,\,\,\,
K_y = \frac{-k_x+k_y}{\sqrt{2}} ,
\end{equation}
we obtain a parabolic band with anisotropic effective masses:
\begin{eqnarray}
\epsilon_{\bf K} &=& -\mu +4tp -\frac{K_x^2}{2m_x}-\frac{K_y^2}{2m_y} ,
\nonumber \\ 
\frac{1}{m_x} &=& 2tp+4J(1-p^2) , \,\,\,\,\,\,\,\, 
\frac{1}{m_y} = 2tp .
\label{disp2}
\end{eqnarray}
The dispersion law $\epsilon_{\bf k}$ is shown in Figs.~\ref{fig:FS1}
and \ref{fig:FS2} (we have used $p\simeq (U/2t)x$, see
Eq.~(\ref{p1})). When $t^2/U^2\ll 
x$, the anisotropy of the effective masses is weak and the Fermi
surface is almost circular. The minima of the LHB lie
at the corners of the Brillouin zone (Fig.~\ref{fig:FS1}). With
decreasing $x$, the Fermi surface becomes elliptical, and
the minima of the LHB move away from the corners of the Brillouin
zone (Fig.~\ref{fig:FS2}).

For $T=0$, the free energy (per site) $F(\mu)$ is given by (see
Appendix \ref{appI}) 
\begin{eqnarray}
F &=& \frac{1}{N} \sum_{\bf k} \theta(-\epsilon_{\bf k}) 
      \epsilon_{\bf k} \nonumber \\ 
  &=& -\mu -J(1-p^2) -\frac{(4tp-\mu)^2}{4\pi}\sqrt{m_xm_y} .
\label{F1}
\end{eqnarray}
The chemical potential is obtained from $n=1-x=-\partial F/\partial
\mu$,\cite{note0}  i.e.
\begin{equation}
\mu = 4tp-\frac{2\pi x}{\sqrt{m_xm_y}} .
\label{mu1}
\end{equation}
From Eqs.~(\ref{F1}) and (\ref{mu1}), we deduce the energy
$E(x)=F+\mu(1-x)$: 
\begin{equation}
E = -4tpx-J(1-p^2)+\frac{\pi x^2}{\sqrt{m_xm_y}} .
\end{equation}
The spiral pitch is obtained by minimizing the energy $E$ at fixed
hole doping $x$, i.e. $\partial E/\partial p=0$. This gives the
following equation for $p$:
\begin{eqnarray}
&& (4tx-2Jp)[4t^2p^2+8tJp(1-p^2)]^{1/2} \nonumber \\ && 
=\frac{\pi x^2}{2}[8t^2p+8tJ(1-3p^2)] . 
\end{eqnarray}
In the limit $x\ll 1$, we find\cite{note1,Zhou95,Zhou94,Arrigoni00}
\begin{equation}
p\simeq \frac{U}{2t}x .
\label{p1}
\end{equation}
At half-filling, the spin configuration is antiferromagnetic ($p=0$). 
As soon as holes are introduced into the system, a spiral phase is
stabilized. Since $p\leq 1$, Eq.~(\ref{p1}) holds only when $x\leq
x_c=2t/U$. When $x\geq x_c$, the minimum of the energy is reached for
the ferromagnetic state ($p=1$). 

From Eqs.~(\ref{mu1}) and (\ref{p1}), we deduce that the
compressibility $\kappa=\partial n/\partial 
\mu\simeq -1/2U$ is negative in the spiral phase. This signals an
instability towards phase separation. The presence
of imaginary 
spin-wave modes at finite momenta (see Sec.~\ref{sec:swm}) suggests
that other types of ground-states could also be stabilized. Zhou and
Schulz have proposed a charge-density wave (coexisting with
some kind of magnetic order). \cite{Zhou95,Zhou94} This possibility
has also been studied in the framework of the Schraiman and Siggia's
model. \cite{Dombre89} The ferromagnetic
phase, which has a positive compressibility $\kappa\simeq 1/4\pi t$,
is found to be stable. 

\begin{figure}
\epsfysize 1.8 cm
\epsffile[180 355 430 430]{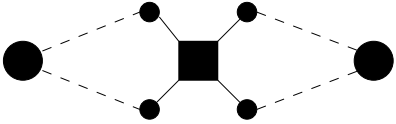}
\caption{Lowest-order correction $F'$ to the free energy $F$ due to
the two-particle vertex $\Gamma^{\rm II}$. } 
\label{fig:Fp}
\end{figure}

Let us now consider the self-energy shown in
Fig.~\ref{fig:SE}b. Although its calculation presents no difficulty,
the determination of its contribution to the free
energy requires to perform a coupling constant integration in order to
ensure a consistent description of the system thermodynamics. Instead of
considering $\Sigma$, we evaluate the lowest order correction $F'$
to the free energy $F$ due to the two-particle vertex $\Gamma^{\rm II}$
(Fig.~\ref{fig:Fp}). This is sufficient to 
show that $\Gamma^{\rm II}$ can be ignored in the low-doping limit
($x\ll 1$). As shown in Appendix \ref{appII}, one finds
\begin{equation}
F'=xJ(1-p^2) .
\end{equation}
$F'$ can be simply taken into account by replacing $J(1-p^2)$ in
Eq.~(\ref{F1}) by $J(1-p^2)(1-x)$. This modification can be ignored
for $x\ll 1$. We conclude that $\Gamma^{\rm II}$ can be ignored in the
low-doping limit. 

The fact that we can neglect the quartic term of the action
$S[\gamma^*,\gamma;{\bf\Omega}]$ justifies the functional integral
formulation of Refs.~\onlinecite{Zhou95,Zhou94}. It is also the reason
why a standard RPA analysis
\cite{Arrigoni91,Dzierzawa92,Zhou95,Zhou94,Brenig96,Kampf96,Arrigoni00}
gives correct results in the strong-coupling limit. Note however that
this conclusion holds only in the {\it low-doping} limit ($x\ll
1$). For an arbitrary doping, the quartic term of 
the action (\ref{action2}) cannot be neglected, and the RPA approach
breaks down since it misses processes of order $O(t/U)$. It should
also be noted that the
UHB Green's function ${\cal G}_\downarrow$ obtained by diagonalizing
the quadratic part of the action (\ref{action2}) (which is the
starting point of the RPA approach) is meaningless in the
strong-coupling regime since it does not correspond to the leading
term within a $t/U$ expansion (see Ref.~\onlinecite{ND} for a detailed
discussion of this point).

\section{Spin-wave modes}
\label{sec:swm}

\subsection{Effective action}

In this section, we derive the effective action for the
fluctuations $\delta{\bf\Omega}_{\bf 
r}={\bf\Omega}_{\bf r}-{\bf\Omega}^{\rm cl}_{\bf r}$ of the spin
variables around their saddle-point value. The calculation is
performed within a perturbative expansion in $t/U$. 

We parametrize the fluctuations  $\delta{\bf\Omega}_{\bf r}$
by \cite{Auerbach} 
\begin{equation}
p_{\bf r} = \frac{1}{2} \delta{\bf\Omega}_{\bf r}\cdot
\hat{\bbox{\theta}}_{\bf r} , \,\,\,\,\,\,
q_{\bf r} = \delta{\bf\Omega}_{\bf r}\cdot\hat{\bbox{\varphi}}_{\bf r} ,
\end{equation}
where $\hat{\bbox{\theta}}_{\bf r}=\hat{\bf z}$ and $\hat{\bbox{\varphi}}_{\bf
r}=(-\sin({\bf Q}\cdot {\bf r}),\cos({\bf Q}\cdot {\bf
r}),0)$. $q_{\bf r}$ ($p_{\bf r}$) corresponds to fluctuations in (out
of) the spiral plane. We easily obtain 
\begin{equation}
p_{\bf r} = \frac{\cos\theta_{\bf r}}{2} , \,\,\,\,\,\,
q_{\bf r} \simeq \varphi_{\bf r}-{\bf Q}\cdot{\bf r} ,
\label{pq}
\end{equation}
where the last equality holds for small fluctuations. The Berry phase
term can be expressed in terms of the variables $p,q$:
\begin{equation}
A^0_{\bf r} = -\frac{i}{2}(p_{\bf r}\dot q_{\bf r}-\dot p_{\bf r}q_{\bf r}).
\end{equation}
Using Eqs.~(\ref{R1}) and (\ref{pq}), we deduce 
\begin{equation}
\delta \hat t_{{\bf rr}'} = \hat t_{{\bf rr}'}-\hat t^{\rm cl}_{{\bf rr}'}
= t_{{\bf rr}'}
\left (
\begin{array}{lr}
A_{{\bf rr}'}  & -B^*_{{\bf rr}'} \\ B_{{\bf rr}'} & A^*_{{\bf rr}'} 
\end{array}
\right )
\end{equation}
where 
\bleq
\begin{eqnarray}
A_{{\bf rr}'} &=& \cos\biggl(\frac{{\bf Q}\cdot({\bf r}-{\bf r}')}{2}\biggr) 
\biggl[ -\frac{(p_{\bf r}-p_{{\bf r}'})^2}{2}
+\frac{i}{2}(p_{\bf r}+p_{{\bf r}'})(q_{\bf r}-q_{{\bf r}'})
-\frac{(q_{\bf r}-q_{{\bf r}'})^2}{8} \biggr] 
\nonumber \\ && 
+ \sin\biggl(\frac{{\bf Q}\cdot({\bf r}-{\bf r}')}{2}\biggr) 
\biggl[ i(p_{\bf r}+p_{{\bf r}'})-\frac{q_{\bf r}-q_{{\bf r}'}}{2}
\biggr] ,
\\ 
B_{{\bf rr}'} &=& \cos\biggl(\frac{{\bf Q}\cdot({\bf r}-{\bf r}')}{2}\biggr) 
\biggl[ p_{\bf r}-p_{{\bf r}'} 
-\frac{i}{2} (q_{\bf r}-q_{{\bf r}'}) \biggr] 
\nonumber \\ && 
+ \sin\biggl(\frac{{\bf Q}\cdot({\bf r}-{\bf r}')}{2}\biggr) 
\biggl[ \frac{i}{2}(p_{\bf r}+p_{{\bf r}'})^2 
- \frac{1}{2}(p_{\bf r}-p_{{\bf r}'})(q_{\bf r}-q_{{\bf r}'})
+ \frac{i}{8}(q_{\bf r}-q_{{\bf r}'})^2 \biggr] 
\end{eqnarray}
\eleq
to quadratic order in $p,q$.  

Taking into account spin fluctuations, the action can be written as
\begin{eqnarray}
S[\gamma^*,\gamma;{\bf\Omega}] &=& S_{\rm cl}[\gamma^*,\gamma] 
-\sum_{{\bf r},{\bf r}'} \int d\tau \gamma^\dagger_{\bf r} \delta \hat
t_{{\bf rr}'} \gamma_{{\bf r}'} \nonumber \\ && 
+ \sum_{{\bf r},\sigma} \sigma \int d\tau A^0_{\bf r} \gamma^*_{{\bf
r}\sigma} \gamma_{{\bf r}\sigma} ,
\end{eqnarray}
where $\gamma_{\bf r}=(\gamma_{{\bf r}\uparrow},\gamma_{{\bf
r}\downarrow})^T$. The action of the spin degrees of freedom is
obtained by integrating out the fermions:
\begin{equation}
e^{-S[p,q]} = \int {\cal D}[\gamma]
e^{-S[\gamma^*,\gamma;{\bf\Omega}]} .
\end{equation}
For small fluctuations around the classical configuration
${\bf\Omega}^{\rm cl}$, it is sufficient to determine $S[p,q]$ to
quadratic order in $p,q$. We find
\begin{equation}
S[p,q] = S_B[p,q] + S_1[p,q] + S_2[p,q] ,
\end{equation}
where
\begin{eqnarray}
S_B &=& \sum_{{\bf r},\sigma} \sigma \int d\tau A^0_{\bf r}
\langle \gamma^*_{{\bf r}\sigma} \gamma_{{\bf r}\sigma} \rangle_{\rm
cl} , \\
S_1 &=& -\sum_{{\bf r},{\bf r}'} \int d\tau \langle\gamma^\dagger_{\bf
r} \delta \hat t_{{\bf rr}'} \gamma_{{\bf r}'} \rangle_{\rm cl} , \\
S_2 &=& -\frac{1}{2} \sum_{{\bf r}_1,{\bf r}'_1,{\bf r}_2,{\bf r}'_2} 
\int d\tau_1 d\tau_2 \langle \gamma^\dagger_{{\bf r}_1}
\delta \hat t_{{\bf r}_1{\bf r}'_1}(\tau_1) \gamma_{{\bf r}'_1}
\nonumber \\ && \times  
\gamma^\dagger_{{\bf r}_2} \delta \hat t_{{\bf r}_2{\bf r}'_2}
(\tau_2) \gamma_{{\bf r}'_2} \rangle_{\rm cl} . \nonumber \\ && 
\end{eqnarray}
$\langle \cdots\rangle_{\rm cl}$ means that the average is taken with the
saddle-point action $S_{\rm cl}[\gamma^*,\gamma]$, only the connected
part being considered. $S_B$ is a Berry phase term, $S_1$ a
first-order cumulant, and $S_2$ a second-order cumulant. Note that
$A^0$ does not contribute to the second-order cumulant, since it is of
second order in $p,q$. 

\subsubsection{Berry phase term $S_B$}

Since $\langle\gamma^*_{{\bf r}\sigma}\gamma_{{\bf
r}\sigma}\rangle_{\rm cl}=(1-x)\delta_{\sigma,\uparrow}$, we obtain
\begin{equation}
S_B = - \frac{i}{2} (1-x) \sum_{\bf r} \int d\tau (p_{\bf r}\dot
q_{\bf r}- \dot p_{\bf r} q_{\bf r}) .
\end{equation}
$S_B$ is the standard expression for the Berry phase term of localized
spins, with a reduction factor $1-x$ due to doping.

\subsubsection{First-order cumulant $S_1$} 

\begin{figure}
\epsfxsize 7. cm
\epsffile[40 485 250 530]{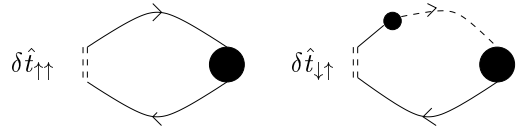}
\caption{Diagrammatic representation of $S_1'$ and $S_1''$.} 
\label{fig:S1}
\end{figure}

We write the first-order cumulant as $S_1=S_1'+S_1''$ where $S_1'$
($S''_1$) is of order $O(1)$ ($O(t/U)$). The corresponding Feynman
diagrams are shown in Fig.~\ref{fig:S1}. We find
\begin{eqnarray}
S_1' &=& - \sum_{{\bf r},{\bf r}'} \int d\tau \delta \hat
t_{{\bf r}\uparrow,{\bf r}'\uparrow} 
{\cal G}_\uparrow({\bf r}'-{\bf r},\tau=0^-) \nonumber \\ 
&=& - \frac{1}{\beta} \sum_{{\bf k},\omega} \delta \hat
t_{{\bf k}\uparrow,{\bf k}\uparrow} 
{\cal G}_\uparrow({\bf k},i\omega) , \\ 
S_1'' &=& - \sum_{{\bf r},{\bf r}',\sigma} \int d\tau \delta \hat
t_{{\bf r}\sigma,{\bf r}'\bar\sigma} 
{\cal G}_{\bar\sigma\sigma}({\bf r}'-{\bf r},\tau=0^-) \nonumber \\ 
&=& - \frac{1}{\beta} \sum_{{\bf k},\omega.\sigma} \delta \hat
t_{{\bf k}\sigma,{\bf k}\bar\sigma} 
{\cal G}_{\bar\sigma\sigma}({\bf k},i\omega) ,
\end{eqnarray}
where we have introduced
\begin{equation}
\delta \hat t_{{\bf k}\sigma,{\bf k}'\sigma'} = \frac{1}{N} \sum_{{\bf
r},{\bf r}'} e^{-i{\bf k}\cdot {\bf r}+i{\bf k}'\cdot {\bf r}'} 
\delta \hat t_{{\bf r}\sigma,{\bf r}'\sigma'} .
\end{equation}
The Green's function ${\cal G}_\uparrow({\bf k},i\omega)$ is given by
Eq.~(\ref{Gup}). To first order in $t/U$, 
\begin{eqnarray}
{\cal G}_{\bar\sigma\sigma}({\bf k},i\omega) &=& 2t \sin\frac{Q}{2}
X_{\bf k} {\cal G}_\uparrow({\bf k},i\omega)
G^{(0)}_\downarrow(i\omega) \nonumber \\ &\simeq& 
- \frac{2t}{U} \sin\frac{Q}{2}
X_{\bf k} {\cal G}_\uparrow({\bf k},i\omega) ,
\end{eqnarray}
where $X_{\bf k}=\sin k_\nu$. 

Performing the sums over the Matsubara frequencies and using  
\begin{eqnarray}
\delta \hat t_{{\bf k}\uparrow,{\bf k}\uparrow} &=&
\frac{4t}{\sqrt{N}} \sin\frac{Q}{2} X_{\bf k} p_{{\bf q}=0} \nonumber
\\  &&
- \frac{2t}{N} \cos\frac{Q}{2} \sum_{\bf q} 
\biggl(|p_{\bf q}|^2 +\frac{|q_{\bf q}|^2}{4} \biggr) 
(Y _{\bf k}-Y_{{\bf k}+{\bf q}}) \nonumber \\ && 
+ i\frac{t}{N} \cos\frac{Q}{2} \sum_{\bf q} p_{\bf q}q_{-\bf q}
(Y_{{\bf k}+{\bf q}} -Y_{{\bf k}-{\bf q}}) , \\ 
\sum_\sigma \delta \hat t_{{\bf k}\sigma,{\bf k}\bar\sigma} &=& 
\frac{4t}{N} \sin\frac{Q}{2}  \sum_{\bf q} \Bigl[ 
|p_{\bf q}|^2(X_{\bf k}+ X_{{\bf k}+{\bf q}}) 
\nonumber \\ && 
+ \frac{|q_{\bf q}|^2}{4} (X_{\bf k}- X_{{\bf k}+{\bf q}}) \Bigr] ,
\end{eqnarray}
we finally obtain
\begin{eqnarray}
S'_1 &=& - \frac{2tp}{N} \sum_{{\bf k},\tilde q} \theta(-\epsilon_{\bf
k})(Y_{\bf k} -Y_{{\bf k}+{\bf q}}) \biggl(|p_{\tilde q}|^2 +\frac{|q_{\tilde
q}|^2}{4} \biggr)  , \nonumber \\
S''_1 &=& \frac{2J(1-p^2)}{N}  \sum_{{\bf k},\tilde q} \theta(-\epsilon_{\bf
k})  X_{\bf k}  \Bigl[ 
|p_{\tilde q}|^2(X_{\bf k}+ X_{{\bf k}+{\bf q}}) 
\nonumber \\ && 
+ \frac{|q_{\tilde q}|^2}{4} (X_{\bf k}- X_{{\bf k}+{\bf q}}) \Bigr] . 
\end{eqnarray}
where $Y_{\bf k}=\cos k_\nu$.
Here we use the notation $\tilde q=({\bf q},i\omega_\nu)$ with
$\omega_\nu$ a bosonic Matsubara frequency. $p_{\tilde q}$ and
$q_{\tilde q}$ are the Fourier transforms of $p_{\bf r}$ and $q_{\bf
r}$ with respect to space and time.

\subsubsection{Second-order cumulant $S_2$} 

\begin{figure}
\epsfxsize 8. cm
\epsffile[55 385 340 625]{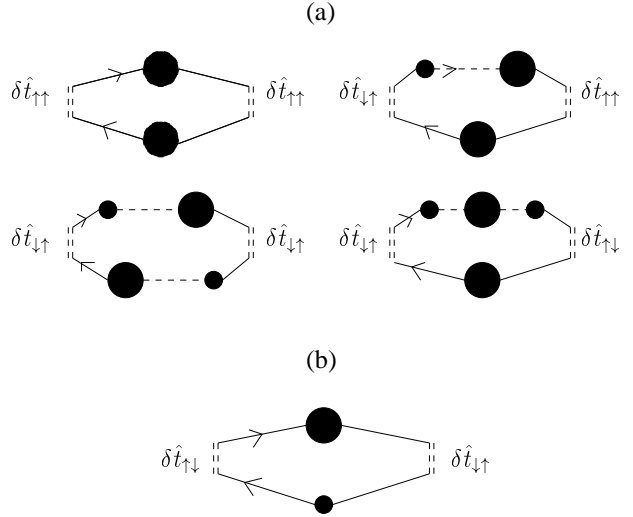}
\caption{Diagrammatic representation of $S_2'$ (a) and $S_2''$ (b).} 
\label{fig:S2}
\end{figure}

\begin{figure}
\epsfysize 3 cm
\epsffile[45 455 260 555]{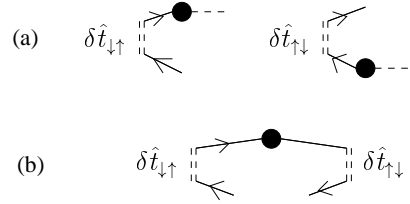}
\caption{
(a) ``Vertex corrections'' generated from the self-energy $\Sigma$
(Fig.~\ref{fig:SE}a) by replacing $\hat t^{\rm cl}_{\sigma\bar\sigma}$
by $\hat t^{\rm cl}_{\sigma\bar\sigma}+\delta \hat
t_{\sigma\bar\sigma}$. (b) Diagram $O(\delta \hat t^2)$ generated by
the same procedure. }
\label{fig:vertex}
\end{figure}

We decompose $S_2$ into two contributions, $S_2'$ and $S_2''$, which
are represented diagrammatically in Fig.~\ref{fig:S2}. $S_2'$ involves
particle-hole excitations in the LHB and describes the dynamical
interaction between holes and spin fluctuations. This interaction can
also be seen as an RKKY-type interaction between localized spins
mediated by the mobile holes.\cite{Arrigoni00} The first diagram of
Fig.~\ref{fig:S2}a gives a contribution which is $O(1)$ in $t/U$. The
other three diagrams take into account ``vertex corrections'' to the
external ``vertices'' $\delta\hat t_{\uparrow\uparrow}$ appearing in
the $O(1)$ contribution. These vertex corrections can be
systematically generated from the self-energy diagram shown in
Fig.~\ref{fig:SE}a. Replacing  $\hat t^{\rm cl}_{\sigma\bar\sigma}$
by $\hat t^{\rm cl}_{\sigma\bar\sigma}+\delta \hat
t_{\sigma\bar\sigma}$ in this diagram, one obtains the vertex
corrections shown in 
Fig.~\ref{fig:vertex}a. Including the latter into the $O(1)$
contribution to $S_2'$, we obtain all the diagrams contributing to
$S_2'$ (Fig.~\ref{fig:S2}a). This procedure also generates a diagram
of order $\delta \hat t^2$ (Fig.~\ref{fig:vertex}b), which gives rise
to $S''_2$ (Fig.~\ref{fig:S2}b). It should be noted here that
$S_2'$ contains diagrams (but not all of them) of order
$O(t^2/U^2)$. We also note that the diagrams contributing to the
first-order cumulant $S_1$ (Fig.~\ref{fig:S1}) can be obtained in the
same way, i.e. by considering the vertex corrections shown in
Fig.~\ref{fig:vertex}a. 

As will be shown below, generating the diagrams for
$S[p,q]$ from the self-energy $\Sigma$ ensures a proper description of the
Goldstone modes. A mere calculation of the action $S[p,q]$ to order 
$O(t/U)$ would yield a contradiction with Goldstone theorem.  

Let us first consider $S_2''$. We find
\begin{eqnarray}
S_2'' &=& \frac{1}{2} \sum_{{\bf r}_1,{\bf r}'_1,{\bf r}_2,{\bf
r}'_2,\sigma} \int d\tau_1 
d\tau_2 \delta \hat t_{{\bf r}_1\sigma,{\bf r}_1'\bar\sigma}(\tau_1) 
\delta \hat t_{{\bf r}_2\bar\sigma,{\bf r}_2'\sigma}(\tau_2) \nonumber
\\ && \times \Bigl[ \delta_{\sigma,\uparrow} 
G^{(0)}_{\downarrow}({\bf r}_1'-{\bf r}_2,\tau_1-\tau_2)  
{\cal G}_\uparrow({\bf r}_2'-{\bf r}_1,\tau_2-\tau_1)  \nonumber \\ && 
+ \delta_{\sigma,\downarrow}
{\cal G}_{\uparrow}({\bf r}_1'-{\bf r}_2,\tau_1-\tau_2)  
G^{(0)}_\downarrow({\bf r}_2'-{\bf r}_1,\tau_2-\tau_1) \Bigr] .  
\label{S2pp1}
\end{eqnarray}
Since spin fluctuations have a characteristic energy scale $J\ll U$,
we can replace $\delta\hat t(\tau_2)$ by $\delta\hat t(\tau_1)$ in
Eq.~(\ref{S2pp1}). This allows to obtain
\begin{eqnarray}
S_2'' &=& - \frac{1}{2U} \sum_{{\bf k},{\bf k}',\sigma} \int d\tau 
\delta \hat t_{{\bf k}'\sigma,{\bf k}\bar\sigma} 
\delta \hat t_{{\bf k}\bar\sigma,{\bf k}'\sigma}  \nonumber \\ &&
\times [ \delta_{\sigma,\downarrow}\theta(-\epsilon_{\bf k}) + 
\delta_{\sigma,\uparrow}\theta(-\epsilon_{{\bf k}'}) ] . 
\end{eqnarray}
To linear order in $p,q$, 
\begin{eqnarray}
\delta \hat t_{{\bf k}\uparrow,{\bf k}'\downarrow} &=&
-\frac{2t}{\sqrt{N}} \cos\frac{Q}{2} (Y_{{\bf k}'}-Y_{\bf k})\Bigl(p_{{\bf
k}-{\bf k}'} + \frac{i}{2} q_{{\bf k}-{\bf k}'}\Bigr) ,  \nonumber \\ 
\delta \hat t_{{\bf k}\downarrow,{\bf k}'\uparrow} &=&
\frac{2t}{\sqrt{N}} \cos\frac{Q}{2} (Y_{{\bf k}'}-Y_{\bf k})\Bigl(p_{{\bf
k}-{\bf k}'} - \frac{i}{2} q_{{\bf k}-{\bf k}'}\Bigr) ,
\end{eqnarray}
so that we finally obtain 
\begin{equation}
S_2'' = -\frac{Jp^2}{N} \sum_{{\bf k},\tilde q} \theta(-\epsilon_{\bf k})
(Y_{{\bf k}+{\bf q}}-Y_{\bf k})^2 \biggl(|p_{\tilde q}|^2 +\frac{|q_{\tilde
q}|^2}{4} \biggr) .
\end{equation}
The sum of $S_1$ and $S_2''$ can be written in the form
\begin{equation}
S_1+S_2''=\sum_{\tilde q} \Bigl[ |p_{\tilde q}|^2(h_{\bf
    q}+\Delta_{\bf q})+\frac{|q_{\tilde q}|^2}{4}(h_{\bf
    q}-\Delta_{\bf q}) \Bigr] ,
\end{equation}
where
\begin{eqnarray}
h_{\bf q} &=& \frac{1}{N} \sum_{\bf k} \theta(-\epsilon_{\bf k}) 
\Bigl[ -2tp(Y_{\bf k}-Y_{{\bf k}+{\bf q}}) \nonumber \\ && 
+2J(1-p^2)X^2_{\bf k}-Jp^2(Y_{{\bf k}+{\bf q}}-Y_{\bf
    k})^2 \Bigr] , \nonumber \\ 
\Delta_{\bf q} &=& \frac{2J(1-p^2)}{N} \sum_{\bf k}
\theta(-\epsilon_{\bf k}) X_{\bf k}X_{{\bf k}+{\bf q}} .
\end{eqnarray}

Let us now consider $S_2'$. It can be written as
\begin{equation}
S_2' = \frac{1}{2} \sum_{{\bf k},\tilde q} 
\delta \hat t^R_{{\bf k}+{\bf q}\uparrow,{\bf k}\uparrow}(i\omega_\nu) 
\delta \hat t^R_{{\bf k}\uparrow,{\bf k}+{\bf q}\uparrow}(-i\omega_\nu) 
\Pi({\bf k};\tilde q) ,
\end{equation}
where
\begin{eqnarray}
\Pi({\bf k};\tilde q) &=& \frac{1}{\beta} \sum_\omega  {\cal
G}_\uparrow({\bf k},i\omega)  {\cal G}_\uparrow({\bf k}+{\bf
q},i\omega+i\omega_\nu) \nonumber \\ &=& \frac{\theta(-\epsilon_{\bf
k})-\theta(-\epsilon_{{\bf k}+{\bf q}})}{\epsilon_{\bf
k}-\epsilon_{{\bf k}+{\bf q}}+i\omega_\nu} 
\end{eqnarray}
is the polarization bubble of the fermions in the LHB. 
$\delta\hat t^R$ is a renormalized vertex which takes care of the
vertex corrections discussed above. To linear order in $p,q$ 
\begin{eqnarray}
\delta\hat t^R_{{\bf k}'\uparrow,{\bf k}\uparrow} &=&
\frac{2t}{\sqrt{N}} \sin\frac{Q}{2} \Bigl[ p_{{\bf k}'-{\bf
k}} (X_{\bf k}\Lambda_{{\bf k},{\bf k}'} + X_{{\bf
k}'}\Lambda_{{\bf k}',{\bf k}} ) \nonumber \\ && + \frac{i}{2} q_{{\bf
k}'-{\bf  k}} (X_{\bf k}\Lambda_{{\bf k},{\bf k}'} - X_{{\bf
k}'}\Lambda_{{\bf k}',{\bf k}} ) \Bigr] ,
\end{eqnarray}
with
\begin{equation} 
\Lambda_{{\bf k},{\bf k}'} = 1-\frac{2t}{U} \cos\frac{Q}{2}(Y_{\bf
k}-Y_{{\bf k}'}). 
\end{equation}
We finally obtain 
\begin{equation}
S_2' = \sum_{\tilde q} (p^*_{\tilde q},q^*_{\tilde q}) 
\left( 
\begin{array}{lr}
S_+(\tilde q) & \frac{i}{2}S_{+-}(\tilde q) \\
-\frac{i}{2}S_{+-}(\tilde q) & \frac{1}{4}S_-(\tilde q) 
\end{array}
\right) 
\left(
\begin{array}{l}
p_{\tilde q} \\ q_{\tilde q} 
\end{array}
\right) ,
\end{equation}
where we have introduced 
\begin{eqnarray}
S_\pm(\tilde q) &=& \frac{1}{2N} \sum_{\bf k} \Pi({\bf k};\tilde q)
\Lambda^2_\pm ({\bf k};{\bf q}) , \\
S_{+-}(\tilde q) &=& \frac{1}{2N} \sum_{\bf k} \Pi({\bf k};\tilde q)
\Lambda_+ ({\bf k};{\bf q})\Lambda_- ({\bf k};{\bf q}) , 
\end{eqnarray}
and
\begin{equation}
\Lambda_\pm ({\bf k};{\bf q}) = 2t \sin\frac{Q}{2} (X_{\bf
k}\Lambda_{{\bf k},{\bf k}+{\bf q}} \pm X_{{\bf k}+{\bf q}}
\Lambda_{{\bf k}+{\bf q},{\bf k}} ) .
\end{equation}

The total action $S=S_B+S_1+S_2$ is then written as
\bleq
\begin{equation}
S = \sum_{\tilde q} (p^*_{\tilde q},q^*_{\tilde q}) 
\left( 
\begin{array}{lr}
h_{\bf q}+\Delta_{\bf q}+S_+(\tilde q) & 
-(1-x)\frac{\omega_\nu}{2}+\frac{i}{2}S_{+-}(\tilde q) \\
(1-x)\frac{\omega_\nu}{2}-\frac{i}{2}S_{+-}(\tilde q) & 
\frac{1}{4}(h_{\bf q}-\Delta_{\bf q}+S_-(\tilde q)) 
\end{array}
\right) 
\left(
\begin{array}{l}
p_{\tilde q} \\ q_{\tilde q} 
\end{array}
\right) .
\end{equation}
The spin-wave excitations are obtained from the equation
\begin{equation}
[h_{\bf q}+\Delta_{\bf q}+S_+({\bf q},\omega)]
[h_{\bf q}-\Delta_{\bf q}+S_-({\bf q},\omega)]
= [(1-x)\omega+S_{+-}({\bf q},\omega)]^2 ,
\label{sw1}
\end{equation}
\eleq
where $S({\bf q},\omega)=S({\bf q},i\omega_\nu\to \omega+i0^+)$ is the
retarded part of $S(\tilde q)$. Eq.~(\ref{sw1}) was first obtained in
Refs.~\onlinecite{Zhou95,Zhou94}.

\subsection{Goldstone modes}

In this section, we show that the equation for the spin-wave
excitations [Eq.~(\ref{sw1})] accounts for Goldstone modes at ${\bf
q}=0$ and ${\bf q}=\pm {\bf Q}$ in agreement with the Goldstone
theorem for a spiral spin arrangement. \cite{Rastelli85,Gan91}

$\Lambda_-({\bf k};{\bf q})$, $S_-({\bf q},\omega)$ and $S_{+-}({\bf
q},\omega)$ vanish at ${\bf q}=0$.
Since $h_{{\bf q}=0}=\Delta_{{\bf q}=0}$, we conclude that
$\omega=0$ is solution for ${\bf q}=0$. This mode satisfies $p_{\bf
r}=0$ and therefore corresponds to a global rotation in the spiral plane.

When ${\bf q}={\bf Q}$, we have the properties
\begin{eqnarray}
h_{\bf Q}+\Delta_{\bf Q} &=& \frac{1}{N} \sum_{\bf k}
\theta(-\epsilon_{\bf k})(\epsilon_{{\bf k}+{\bf Q}}-\epsilon_{\bf k}),
\nonumber \\ 
S_+({\bf Q},\omega=0) &=& - \frac{1}{N} \sum_{\bf k}
\theta(-\epsilon_{\bf k})(\epsilon_{{\bf k}+{\bf Q}}-\epsilon_{\bf k}).
\end{eqnarray}
The last equality follows from $\Lambda_+({\bf k};{\bf Q}) =
\epsilon_{{\bf k}+{\bf Q}}-\epsilon_{\bf k}$. We have also used the
fact that $S_+({\bf Q},\omega=0)$ is purely real since particle-hole
excitations are gapped for ${\bf q}={\bf Q}$ at low doping (see
Fig.~\ref{fig:swdiag}). Using $\Lambda_\pm(-{\bf 
k}-{\bf q};{\bf q})=\mp \Lambda_\pm({\bf k};{\bf q})$, one can also
show that $S_{+-}({\bf q},\omega=0)=0$. This result, together with 
$h_{\bf Q}+\Delta_{\bf Q}+S_+({\bf Q},\omega=0)=0$, ensures the
existence of a gapless mode for ${\bf q}={\bf Q}$. A similar reasoning
holds for ${\bf q}=-{\bf Q}$. Goldstone modes at $\pm {\bf Q}$ satisfy
$q_{\bf r}=0$ and correspond to fluctuations out of the spiral plane.

\subsection{Analytical expression of the spin-wave excitations}

The long-wavelength spin-wave modes have been studied in detail
before. \cite{Gan91,Zhou95,Zhou94,Brenig96,Kampf96} They are observed
only for very low doping. Above a certain hole concentration, these modes
completely dissolve into the particle-hole excitation spectrum. From
now on, we focus on the region $q\gg K_F(\alpha)$, where
$K_F(\alpha)=O(\sqrt{x})$ is the ``Fermi wave vector'' in the direction
specified by the angle $\alpha$ (see Appendix \ref{appI}). In this
momentum range, it is possible to derive an analytical expression of
the spin-wave excitations. 

\begin{figure}
\epsfysize 12 cm
\epsffile[55 270 290 625]{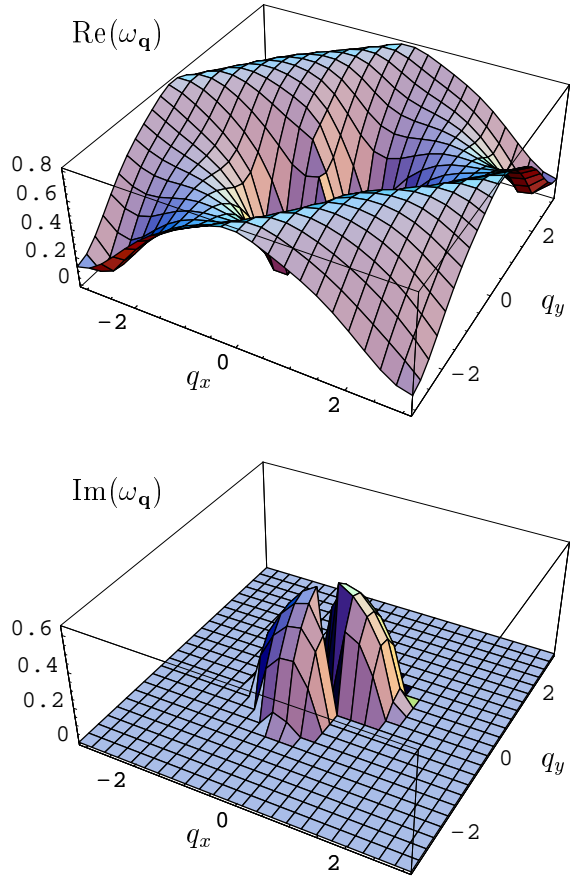}
\caption{
Real and imaginary parts of the spin-wave energy $\omega_{\bf q}$
($U/t=10$ and $x=0.04$).   }
\label{fig:sw}
\end{figure}

\begin{figure}
\epsfxsize 8 cm
\epsffile[40 355 300 515]{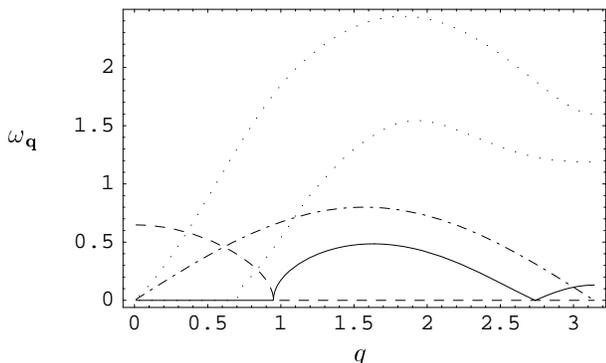}
\caption{
Spin-wave energy in the direction ${\bf q}=(q,q)$ for $U/t=10$ and
$x=0.04$. Solid line: ${\rm Re}(\omega_{\bf q})$ (note the Goldstone
mode at $q=Q \simeq 2.74$), dashed line: ${\rm Im}(\omega_{\bf q})$,
dot-dashed line: excitation spectrum at half-filling
[Eq.~(\ref{swhf})]. The dotted lines indicate the boundaries of the
particle-hole excitation spectrum.  }
\label{fig:swdiag}
\end{figure}

As shown below, the spiral phase sustains both real and imaginary
modes. The latter signal an instability of the spiral phase. However,
the (short-wavelength) modes with a real excitation energy are
expected to survive in the actual ground-state of the system. Since
they lie below the particle-hole excitation continuum (see
Fig.~\ref{fig:swdiag}), the imaginary part of $S_\pm({\bf q},\omega)$
and $S_{+-}({\bf q},\omega)$ does not need to be
considered. Furthermore, $\omega$ can be neglected against
$\epsilon_{{\bf k}+{\bf q}}-\epsilon_{\bf k}$ in the calculation of
$S_\pm$ and $S_{+-}$.  Since
$S_{+-}({\bf q},\omega=0)=0$, the equation for the spin-wave modes
$\omega_{\bf q}$ then reduces to
\begin{equation}
(1-x)^2 \omega^2_{\bf q} = [h_{\bf q}+\Delta_{\bf q}+S_+({\bf q})]
 [h_{\bf q}-\Delta_{\bf q}+S_-({\bf q})],
\label{sw2}
\end{equation}
where $S_\pm({\bf q})=S_\pm({\bf q},\omega=0)$. The summations over
${\bf k}$ involved in $h_{\bf q}$, $\Delta_{\bf q}$ and $S_\pm({\bf q})$ 
are easily carried out in the low-doping limit. For an arbitrary function
$f({\bf k})$, we have
\begin{eqnarray}
\frac{1}{N}\sum_{\bf k} \theta(-\epsilon_{\bf k})f({\bf k}) &=& 
\frac{1}{N}\sum_{\bf k}(1-\theta(\epsilon_{\bf k}))f({\bf k}) \nonumber \\ 
&\simeq& \frac{1}{N} \sum_{\bf k} f({\bf k}) - x f({\bf k}=0) ,
\label{sum1}
\end{eqnarray}
where we have used $(1/N)\sum_{\bf k}\theta(\epsilon_{\bf k})=x$. In
general, the sum over the entire Brillouin zone is easily
evaluated. Eq.~(\ref{sum1}) gives the leading contribution for $x\ll
1$. Thus we find
\begin{eqnarray}
h_{\bf q} &=& 2J(1-p^2)-Jp^2(2-Y_{\bf q}) \nonumber \\ && 
+x[2tp(2-Y_{\bf q})+Jp^2(2-Y_{\bf q})^2] , \label{hq1} \\
&=& 2J(1-p^2)+xJp^2(2-Y_{\bf q})^2 , \label{hq2} \\
\Delta_{\bf q} &=& J(1-p^2)Y_{\bf q} , \label{Dq}\\
S({\bf q}) &=& x4t^2(1-p^2) \frac{X^2_{\bf q}}{\epsilon_{\bf
q}-\epsilon_{{\bf q}=0}} \Bigl[1-\frac{2tp}{U}(Y_{\bf q}-2)\Bigr]^2 , 
\label{Sq}
\end{eqnarray}
where $S({\bf q})=S_+({\bf q})=S_-({\bf q})$. Eq.~(\ref{hq2}) has
been obtained using $p=(U/2t)x$ and is therefore not valid in the
ferromagnetic phase ($x> x_c$). From Eq.~(\ref{sw2}), we deduce
\bleq
\begin{equation}
\omega_{\bf q} = \frac{J}{1-x} \Biggl\lbrace \Biggl[ 2
(1-p^2) + xp^2(2-Y_{\bf q})^2 + (1-p^2) \frac{X^2_{\bf q}}{Y_{\bf
    q}-2} \frac{\bigl(1-x(Y_{\bf q}-2)\bigr)^2}{1-x \frac{1-p^2}{p^2}
  \frac{X^2_{\bf q}}{Y_{\bf q}-2}} \Biggr]^2 
-\bigl[ (1-p^2)Y_{\bf q} \bigr]^2 \Biggr\rbrace ^{1/2} .
\label{sw3}
\end{equation}
\eleq
At half-filling, Eq.~(\ref{sw3}) reproduces the spin-wave
excitations of the Heisenberg antiferromagnet 
\begin{equation}
\omega_{\bf q} = 2J\Biggl[1- \biggl(\frac{\cos q_x+\cos q_y}{2}
\biggr)^2 \Biggr]^{1/2} . 
\label{swhf}
\end{equation}
Thus Eq.~(\ref{sw3}) turns out to be correct over the entire Brillouin
zone when $x=0$. However, away from half-filling, it is not valid in
the vicinity of ${\bf q}=0$ since it does not describe the
long-wavelength Goldstone mode. 
Using $(1-p^2)X^2_{\pm{\bf Q}}=p^2(Y_{\pm{\bf Q}}-2)^2$, one easily
verifies that $\omega_{\pm{\bf Q}}=0$. Thus the analytical result
(\ref{sw3}) satisfies the Goldstone theorem. 
From Eq.~(\ref{sw3}), we deduce that $\omega_{\bf q}$ is either real
or purely imaginary. Figs.~\ref{fig:sw} and \ref{fig:swdiag} shows the
real and imaginary parts of $\omega_{\bf q}$. We obtain a very good
agreement with the numerical results of
Refs.~\onlinecite{Zhou95,Zhou94,Brenig96,Kampf96}.  

\begin{figure}
\epsfxsize 8.3 cm
\epsffile[25 355 300 515]{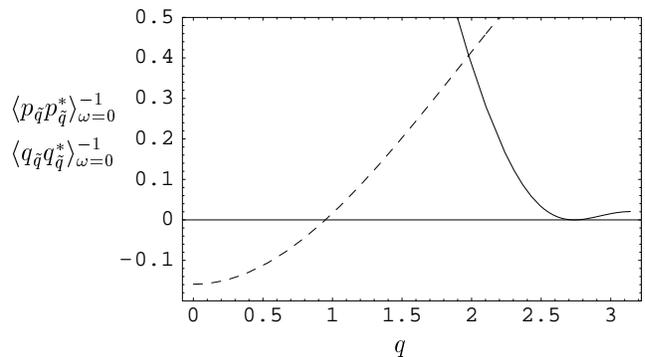}
\caption{ 
Inverse propagators $\langle p_{\tilde q}p^*_{\tilde
q}\rangle_{\omega=0}^{-1}$ (solid line) and $\langle q_{\tilde q}q^*_{\tilde
q}\rangle_{\omega=0}^{-1}$ (dashed line) in the direction ${\bf q}=(q,q)$. 
 }
\label{fig:ppqq}
\end{figure}

The existence of complex frequencies
$\omega_{\bf q}$ in some regions of the Brillouin zone signals an
instability of the spiral phase. The origin of this instability can be
traced back by considering the static correlation functions (with the
analytic continuation $i\omega_\nu\to\omega$) 
\begin{eqnarray}
\langle p_{\tilde q}p^*_{\tilde q}\rangle_{\omega=0} &=&
\frac{1}{2(h_{\bf q}+\Delta_{\bf q}+S({\bf q}))} , \nonumber \\
\langle q_{\tilde q}q^*_{\tilde q}\rangle_{\omega=0} &=&
\frac{2}{h_{\bf q}-\Delta_{\bf q}+S({\bf q})} ,
\label{ppqq}
\end{eqnarray}
where $h_{\bf q}$, $\Delta_{\bf q}$ and $S({\bf q})$ are defined by
Eqs.~(\ref{hq2}-\ref{Sq}). While 
$\langle p_{\tilde q}p^*_{\tilde q}\rangle^{-1}_{\omega=0}$ is always
positive, $\langle q_{\tilde q}q^*_{\tilde q}\rangle^{-1}_{\omega=0}$
turns out to be negative where the excitation energy $\omega_{\bf q}$
becomes imaginary (see Fig.~\ref{fig:ppqq}). [Compare Eqs.~(\ref{sw2})
and (\ref{ppqq}).] This implies a negative spin stiffness (and
therefore an instability) for spin 
fluctuations {\it in} the spiral plane.\cite{Zhou95,Zhou94,Kampf96} As
shown in Refs.~\onlinecite{Zhou95,Zhou94}, the imaginary modes also
lead to a negative charge susceptibility. This suggests that the
actual ground-state could exhibit a charge-density wave (coexisting
with spin order). Contrary to
the phase separated state, the charge-density wave could survive the
presence of long-range Coulomb interaction. 

When $x\gg t^2/U^2$ (i.e. $x/p^2\ll 1$), it is tempting to neglect
altogether the $x$ dependent terms in Eq.~(\ref{sw3}). This leads to 
\begin{equation}
\omega_{\bf q} = \frac{J(1-p^2)}{1-x} \Biggl\lbrace \biggl[ 2+
\frac{X^2_{\bf q}}{Y_{\bf q}-2}\biggr]^2-Y^2_{\bf q}
\Biggr\rbrace^{1/2},
\label{as1}
\end{equation}
which is the result obtained by Arrigoni and Strinati within an RPA
analysis in the regime $t^2/U^2\ll x\ll 1$. \cite{Arrigoni00}
Although Eq.~(\ref{as1}) reproduces the main features of the
spin-wave-excitation spectrum (see Fig.~1 in
Ref.~\onlinecite{Arrigoni00}), it incorrectly predicts a
vanishing of $\omega_{\bf q}$ along the line $q_x=q_y$. Even if there
is no contradiction with Goldstone theorem, the approximation leading
to Eq.~(\ref{as1}) is too crude to yield a correct description of the
Goldstone modes at $\pm{\bf Q}$.

\subsection{Spin-wave modes in the ferromagnetic phase} 

When $x\geq x_c$, the ground-state is ferromagnetic with $p=1$,
i.e. ${\bf Q}=(2\pi,2\pi)\equiv (0,0)$. From Eqs.~(\ref{hq1}),
(\ref{Dq}) and (\ref{Sq}), we obtain the spin-wave excitations
\begin{equation}
(1-x) \omega_{\bf q} = 2t (x-x_c)(2-Y_{\bf q})+xJ(2-Y_{\bf q})^2.
\label{sw4}
\end{equation}
Since $Y_{{\bf q}=0}=2$, the mode at ${\bf q}=0$ is gapless, as
expected for a ferromagnetic spin arrangement. For ${\bf q}\to 0$, 
\begin{equation}
\omega_{\bf q} \simeq \frac{J_{\rm eff}}{2}  q^2, \,\,\,\,\,\,
J_{\rm eff}=2t \frac{x-x_c}{1-x} .
\end{equation}
The instability of the ferromagnetic phase against the formation of a
spiral phase is signaled by a softening of the
spin-wave modes, since the effective ferromagnetic exchange constant
$J_{\rm eff}$ vanishes when $x\to x_c$. Note that for $x=x_c$,
Eqs.~(\ref{sw3}) and (\ref{sw4}) coincide and give $\omega_{\bf
q}=xJ(2-Y_{\bf q})^2/(1-x)$.

\section{Conclusion}

In this paper, we have analyzed the semiclassical limit of the 2D
Hubbard model in the framework of the spin-particle-hole
coherent-state path integral. The main characteristic of the latter
lies in the clear distinction between charge ($\gamma$) and spin
(${\bf\Omega}$) degrees of freedom. The semiclassical analysis
consists in expanding around a broken-symmetry ground-state by making
a saddle-point approximation on the spin variables, while the
fermionic degrees of freedom are integrated out within a systematic
$t/U$ expansion. This should be contrasted with the Hartree-Fock/RPA
theory whose validity in the strong-coupling limit is not obvious even
at the semiclassical level. Nevertheless, we have shown that in the
low-doping limit, the Hartree-Fock/RPA theory does recover the correct
$t/U$ expansion. We have also justified the strong-coupling functional
integral approach of Refs.~\onlinecite{Zhou95,Zhou94} where the
electron-electron interaction is considered within a large-$U$
Hartree-Fock approximation while the SU(2) spin-rotation invariance is
maintained by introducing a fluctuating spin-quantization axis. We
expect however these approaches to break down with increasing doping
since they do not capture all processes of order $t/U$. In the
spin-particle-hole coherent-state path integral, a correct treatment
of all processes of order $t/U$ follows from the consideration of
the quartic term of the action $S[\gamma^*,\gamma;{\bf\Omega}]$
[Eq.~(\ref{action1})].\cite{note2} 

The spin-particle-hole coherent-state path integral provides a
convenient framework for studying spin-wave modes around the
semiclassical ground-state. In particular, it allows a systematic spin
conserving analysis which ensures a proper description of the
Goldstone modes of the system. From the mean-field fermionic
self-energy, one can generate all vertex corrections to be taken into
account in the propagator of the spin-wave modes. A mere calculation
to order $O(t/U)$ 
of the effective action $S[p,q]$ of spin fluctuations would give a
contradiction with Goldstone theorem. Given this observation, it is
not surprising that a $t/U$ expansion of the RPA equations for the
collective modes in the spiral phase meets with difficulties regarding
a correct description of the Goldstone modes. \cite{Arrigoni00} 

Our main result is an analytical expression of the spin-wave
excitations of the spiral phase over the entire Brillouin zone except
in the vicinity of ${\bf q}=0$. Besides the gapless mode at ${\bf
q}=0$, we find two Goldstone modes at $\pm {\bf Q}$ as expected for a
spiral spin order. Our result agrees with previous numerical
calculations and improves the analytical expression obtained in
Ref.~\onlinecite{Arrigoni00}. Although the spiral phase is unstable,
as shown by a negative mean-field susceptibility and the presence of
imaginary modes, we expect the short-wavelength fluctuation modes
(with real energies) to survive in the actual ground-state of the
system.

\bleq

\appendix

\section{Free energy in the mean-field state}
\label{appI}
In order to calculate the free energy $F$ [Eq.~(\ref{F1})], it is
convenient to introduce a ``Fermi wave vector'' ${\bf
K}_F(\alpha)=K_F(\alpha)(\cos\alpha,\sin\alpha)$ defined by
$\epsilon_{{\bf K}_F(\alpha)}=0$. From Eq.~(\ref{disp2}), we deduce
\begin{equation}
K_F^2(\alpha) = 2 
\frac{4tp-\mu}{\frac{\cos^2\alpha}{m_x}+\frac{\sin^2\alpha}{m_y}} .
\end{equation}
The free energy can be written as $F = N^{-1} \sum_{\bf k}
(1-\theta(\epsilon_{\bf k}))\epsilon_{\bf k}$ where 
\begin{eqnarray}
\frac{1}{N} \sum_{\bf k}\epsilon_{\bf k} &=& -\mu-J(1-p^2) , \label{A3}
\\
\frac{1}{N} \sum_{\bf k}\theta(\epsilon_{\bf k}) \epsilon_{\bf k} &=& 
\frac{1}{N} \sum_{\bf K}\theta(\epsilon_{\bf K}) \epsilon_{\bf K}
\nonumber \\ &\simeq& 
\int_0^{2\pi} \frac{d\alpha}{2\pi} \int_0^{K_F(\alpha)}
\frac{dK}{2\pi} K \Biggl[ -\mu+4tp-\frac{K^2}{2}\Biggl
( \frac{\cos^2\alpha}{m_x} + \frac{\sin^2\alpha}{m_y} \Biggr) \Biggr]
\nonumber \\ &=& \frac{(4tp-\mu)^2}{4\pi} \sqrt{m_xm_y} \label{A4}. 
\end{eqnarray}
Eq.~(\ref{F1}) follows from (\ref{A3}) and (\ref{A4}).

\section{Correction $F'$ to the free energy}
\label{appII}

The contribution $F'$ to the free energy (Fig.~\ref{fig:Fp}) is given by
\begin{equation}
F' = \frac{1}{N^2\beta^2} \sum_{{\bf k},{\bf k}'}
\sum_{\omega,\omega'} 
\Gamma^{\rm II}_{\uparrow\downarrow,\uparrow\downarrow}
(i\omega,i\omega';i\omega,i\omega') 
[G^{(0)}_\uparrow(i\omega)]^2 {\cal G}_\uparrow({\bf
k},i\omega)(2tp\cos k_\nu)^2 
[G^{(0)}_\downarrow(i\omega')]^2 {\cal G}_\uparrow({\bf
k}',i\omega')\Bigl(2t\sin\frac{Q}{2}\sin k'_\nu\Bigr)^2 .
\end{equation}
Performing the sums over $\omega$ and $\omega'$ and retaining the
leading order term in $t/U$, we find (for $T\to 0$) 
\begin{eqnarray}
F' &=& \frac{1}{UN^2} \sum_{{\bf k},{\bf k}'}
\theta(\epsilon_{\bf k})(1-\theta(\epsilon_{{\bf k}'}))
(2tp\cos k_\nu)^2 \Bigl(2t\sin\frac{Q}{2}\sin k'_\nu\Bigr)^2 
\frac{1}{(\epsilon_{\bf k}+\mu)^2} .
\end{eqnarray}
Using
\begin{equation}
\frac{1}{N} \sum_{\bf k} \theta(\epsilon_{\bf k}) 4t^2p^2
\frac{(\cos k_\nu)^2}{(\epsilon_{\bf k}+\mu)^2} \simeq 
\frac{1}{N} \sum_{\bf k} \theta(\epsilon_{\bf k}) 
\frac{(4tp)^2}{(\epsilon_{{\bf k}=0}+\mu)^2} = x
\end{equation}
we obtain $F'= xJ(1-p^2)$ in the limit $x\ll 1$.

\eleq

{}

\ecols 


\begin{thebibliography}{}
\bibitem{note1} More precisely, we find $p=(U/2t)x-(\pi/2)x^{3/2}$ for
$x\ll t^2/U^2$, and $p=(U/2t)x-(\pi U/4t)x^2$ for $x\gg t^2/U^2$.
\bibitem{note2} By integrating out the UHB in the partition function
[see Eq.~(\ref{action1})], one recovers the action of the $t$-$J$
model in the spin-hole coherent state path integral. \cite{Dupuis00}
The two-particle atomic vertex $\Gamma^{\rm II}$ plays a crucial role
in this derivation whatever the value of the doping. At half-filling,
the action of the Heisenberg can be obtained in a more direct way by
integrating out all fermionic degrees of freedom at once. In that case
$\Gamma^{\rm II}$ does not enter the derivation. The role of
$\Gamma^{\rm II}$ is therefore dependent on the way charge or spin
degrees of freedom are integrated out.
\end{thebibliography}
\end{document}